\documentclass[12pt]{article}

\usepackage[utf8]{inputenc}
\usepackage{amsmath, amssymb, amsthm, mathtools, bbm, graphicx, fullpage, caption, subcaption, multirow, url, authblk, hyperref}
\usepackage[dvipsnames]{xcolor}
\usepackage[normalem]{ulem}
\usepackage{rotating}
\usepackage{setspace}
\usepackage{booktabs}
\usepackage{booktabs}   
\usepackage{tabularx}   
\usepackage{amsmath}    
\usepackage{natbib}
\usepackage{lineno}
\title{Circadian Derived Features for Early Discrimination Across Insomnia Severity Levels: At Least 8 Weeks of Monitoring Are Needed for Clinically Meaningful Assessment}

\author[1]{\small{Soheil Saghafi}}
\author[2]{\small{Thomas C. Neylan}}
\author[1]{\small{Qiao Li}}
\author[3,4,5]{\small{Samuel A. McLean}}
\author[1, 6, 7, 8]{\small{Gari D. Clifford}}

\affil[1]{\small{Department of Biomedical Informatics, Emory University}}
\affil[2]{\small{Departments of Psychiatry and Neurology, University of California San Francisco}}
\affil[3]{\small{Institute for Trauma Recovery, University of North Carolina at Chapel Hill}}
\affil[4]{\small{Department of Psychiatry, University of North Carolina at Chapel Hill}}
\affil[5]{\small{Department of Emergency Medicine, University of North Carolina at Chapel Hill}}
\affil[6]{\small{Department of Biomedical Engineering, Georgia Institute of Technology \& Emory University}}
\affil[7]{\small{Department of Biomedical Engineering, Johns Hopkins University}}
\affil[8]{\small{Department of Anesthesia \& Critical Care Medicine, Johns Hopkins University}}

\date{}

\begin{document}

\maketitle

%
%

\begin{abstract}

\textbf{Background:} Wearable devices provide continuous, objective measures of daily activity and offer promise for assessing sleep disorders. However, the minimum monitoring duration needed to differentiate insomnia severity remains unclear. We investigated when wearable-derived behavioral features become informative for distinguishing Insomnia Severity Index (ISI) categories and examined the contribution of Activity Count (AC) and circadian-derived features.

\textbf{Methods:} We analyzed wearable data from 2,305 participants in the Advancing Understanding of Recovery after Trauma (AURORA) study. Separate binary classifiers were developed for four ISI categories across six follow-up periods using AC and circadian-derived feature sets. Performance was evaluated using accuracy, F1-score, precision, recall, and AUROC.

\textbf{Results:} Classification improved with longer monitoring. Across ISI categories, approximately eight weeks was the earliest time point at which wearable-derived features consistently achieved informative discrimination (AUROC $\approx 0.60$), with modest improvements thereafter. Participants without clinically significant insomnia were easiest to identify, reaching an AUROC of 0.693. Circadian-derived features performed comparably to, and in several cases better than, AC features, suggesting that the temporal organization of daily activity provides information beyond overall activity volume.

\textbf{Conclusions:} Approximately eight weeks of longitudinal wearable monitoring may represent a practical minimum for differentiating ISI-defined insomnia categories. Longer monitoring provided only incremental improvements. Circadian behavioral features show promise as digital biomarkers for objective insomnia assessment.

\end{abstract}

\section{Introduction}
Insomnia is one of the most prevalent sleep disorders worldwide, affecting approximately 10--20\% of adults and imposing a substantial burden on both individuals and healthcare systems \cite{Ohayon2002, vanStraten2025, Morin2022, Morin2012}. Characterized by persistent difficulty initiating or maintaining sleep despite adequate opportunity for sleep, insomnia is associated with impaired daytime functioning, reduced quality of life, diminished cognitive performance, and increased healthcare utilization \cite{roth2007insomnia,Baglioni2016}. Chronic insomnia has also been linked to an increased risk of depression, anxiety disorders, cardiovascular disease, metabolic dysfunction, and all-cause mortality, emphasizing its importance as a major public health concern \cite{Baglioni2011,Lovato2019,Morin2012}. In both clinical practice and research, insomnia severity is most commonly assessed using the Insomnia Severity Index, a validated patient-reported questionnaire widely used for evaluating symptom severity and monitoring treatment response \cite{Bastien2001,Morin2011}.

Sleep disturbances are particularly common following traumatic events, where alterations in sleep architecture often emerge during the acute recovery period and may persist for months, contributing to the development of chronic insomnia and other psychiatric disorders such as post-traumatic stress disorder (PTSD) \cite{Germain2013,McLean2019}. Because sleep patterns evolve over time, longitudinal monitoring is essential for characterizing symptom progression and identifying individuals who develop persistent insomnia. Objective behavioral measurements collected throughout recovery therefore have the potential to improve both clinical assessment and long-term patient management.

Recent advances in wearable sensing technologies have created unprecedented opportunities for objective, continuous, and unobtrusive monitoring of human behavior in free-living environments. Compared with laboratory-based polysomnography, wearable devices enable long-term data collection at relatively low cost while minimizing participant burden, making them increasingly attractive for digital health applications, remote patient monitoring, and large-scale longitudinal studies \cite{DEZAMBOTTI2019,Depner2019,Chinoy2020}. Among wearable sensing modalities, wrist-worn accelerometers have become one of the most widely adopted approaches for estimating sleep--wake behavior through actigraphy \cite{AncoliIsrael2003,Martin2011}. Recent machine learning studies have further demonstrated the potential of wearable-derived behavioral features for detecting sleep disorders, estimating sleep quality, and supporting objective assessment of insomnia and related clinical outcomes \cite{Baron2018,Spina2023}.

Wearable activity recordings capture multiple dimensions of human behavior. Conventional actigraphy analyses primarily summarize movement using Activity Count, a measure that reflects the magnitude of physical activity over time \cite{AncoliIsrael2003,Martin2011}. Although AC provides valuable information regarding overall movement intensity, it does not explicitly characterize the temporal organization of daily behavior. In contrast, circadian rhythm analysis quantifies behavioral characteristics such as rhythmicity, stability, regularity, phase, and amplitude of the rest--activity cycle through metrics that capture day-to-day consistency and within-day fragmentation of activity patterns \cite{Witting1990,vanSomeren1999}. These characteristics are particularly relevant because disruption of circadian rhythms is widely recognized as an important physiological mechanism contributing to insomnia and other sleep disorders \cite{Zee2007}. Consequently, circadian-derived behavioral features have the potential to complement conventional activity summaries by characterizing the temporal organization of behavior and capturing alterations in sleep--wake regulation that may not be reflected by activity volume alone.


Despite significant advances in the application of wearable technologies for sleep monitoring and insomnia assessment, an important practical question remains insufficiently explored: \emph{How much longitudinal wearable monitoring is required before behavioral patterns become sufficiently informative to reliably differentiate insomnia severity?} Previous studies have primarily focused on developing and validating predictive models using predefined observation windows or fixed monitoring durations, demonstrating the feasibility of wearable-based sleep assessment without systematically investigating how diagnostic performance evolves as progressively longer periods of longitudinal data become available \cite{DEZAMBOTTI2019,Depner2019,Chinoy2020}. Consequently, although wearable-derived behavioral features have shown considerable promise for characterizing sleep disturbances, clinicians and researchers currently lack evidence-based guidance regarding the minimum duration of wearable monitoring required to obtain reliable and clinically meaningful assessments of insomnia severity. Addressing this knowledge gap has important implications for optimizing remote patient monitoring, reducing participant burden, informing the design of longitudinal studies, and facilitating the practical implementation of wearable-based digital health systems.


In this study, we sought to determine the minimum duration of longitudinal wearable monitoring required to reliably differentiate clinically meaningful levels of insomnia severity. To achieve this objective, we analyzed longitudinal wearable data from 2,305 participants enrolled in the AURORA study \cite{McLean2019}. Wearable recordings collected across six follow-up periods, spanning baseline through 12 months after trauma, were used to develop separate binary classification models for each Insomnia Severity Index (ISI) category. We evaluated how classification performance evolved as progressively longer durations of wearable monitoring became available and compared conventional Activity Count (AC) features with circadian-derived behavioral features to determine whether the temporal organization of daily activity provides complementary information beyond overall activity magnitude. By identifying the earliest monitoring duration at which wearable-derived behavioral features become sufficiently informative for reliable differentiation of ISI-defined insomnia categories, this work provides practical guidance for the design and implementation of wearable-based insomnia assessment in both clinical practice and longitudinal research.


This study provides a systematic evaluation of how the duration of longitudinal wearable monitoring influences the ability to differentiate clinically meaningful levels of insomnia severity. By examining wearable recordings collected across six follow-up periods, we identify the earliest monitoring duration at which wearable-derived behavioral features become sufficiently informative for reliable classification of ISI-defined insomnia categories. In addition, we compare conventional Activity Count features with circadian-derived behavioral features to assess whether the temporal organization of daily activity provides complementary information beyond overall activity magnitude. Together, these findings provide practical guidance for optimizing wearable-based insomnia assessment, reducing unnecessary monitoring burden, and informing the design of future longitudinal digital health studies.

\subsection{Dataset and Preprocessing}
\label{sec:data}

\begin{figure}[t]
\centering

\begin{minipage}[c]{0.42\textwidth}
    \centering
    \includegraphics[width=\linewidth]{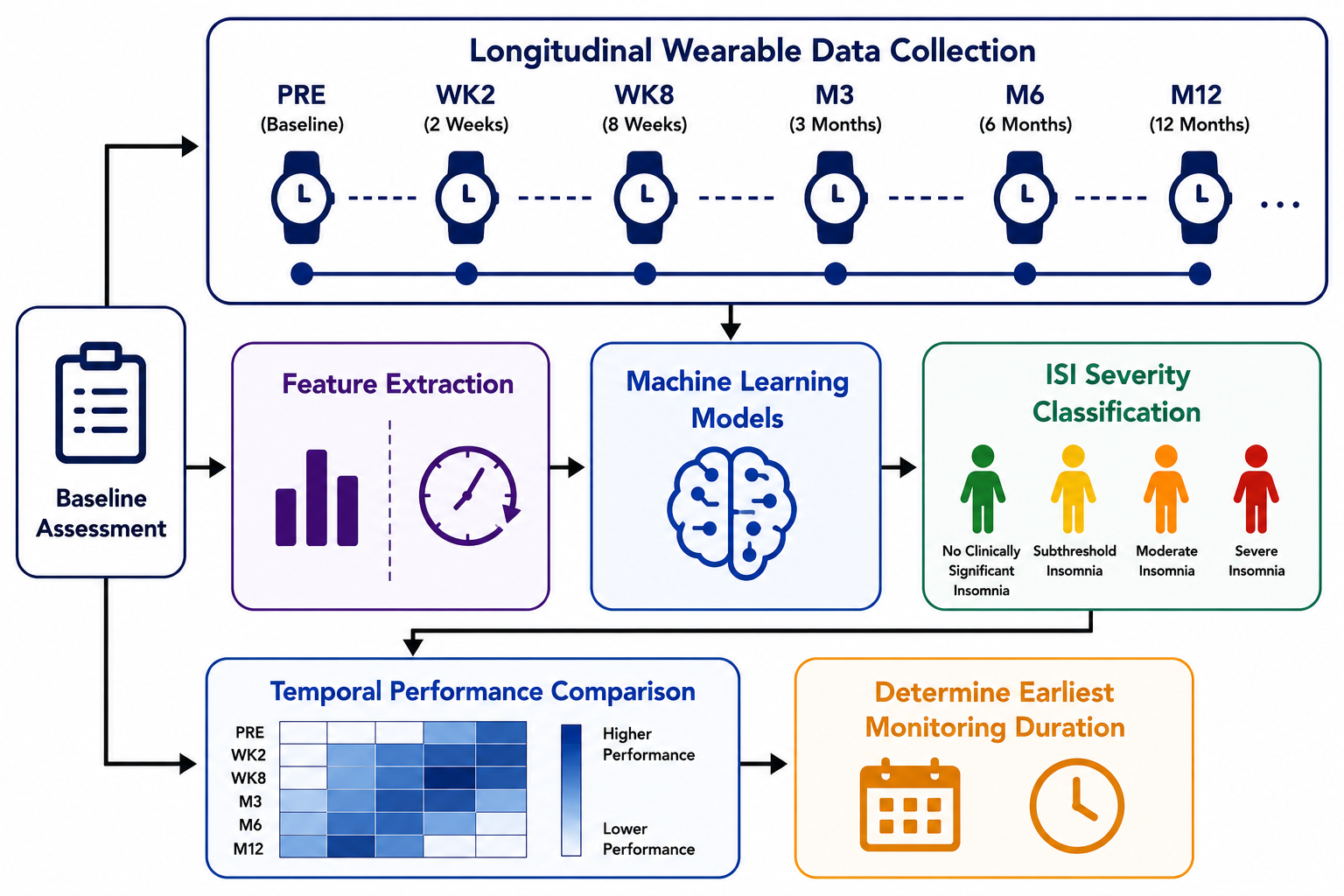}
\end{minipage}
\hfill
\begin{minipage}[c]{0.54\textwidth}
    \caption{\textbf{Schematic of the data collection process.}
    At study enrollment, participants aged 18–75 were instructed to wear a research-grade wristwatch (Verily Life Sciences) for at least 21 hours per day during the first several weeks of follow-up.}
    \label{fig:Fig_Trauma}
\end{minipage}

\end{figure}

This study utilizes longitudinal actigraphy data collected as part of the AURORA study, a large, prospective, multicenter cohort designed to investigate recovery trajectories following traumatic events \cite{McLean2019}. The AURORA study enrolled adults aged 18--75 years who presented to participating emergency departments following trauma exposure. As illustrated in Figure~\ref{fig:Fig_Trauma}, participants were provided with a research-grade wrist-worn wearable device (Verily Life Sciences) and instructed to wear it continuously throughout the longitudinal follow-up period, enabling objective and unobtrusive monitoring of daily behavioral activity under free-living conditions.

\begin{figure*}
\centering
\includegraphics[width=0.9\textwidth]{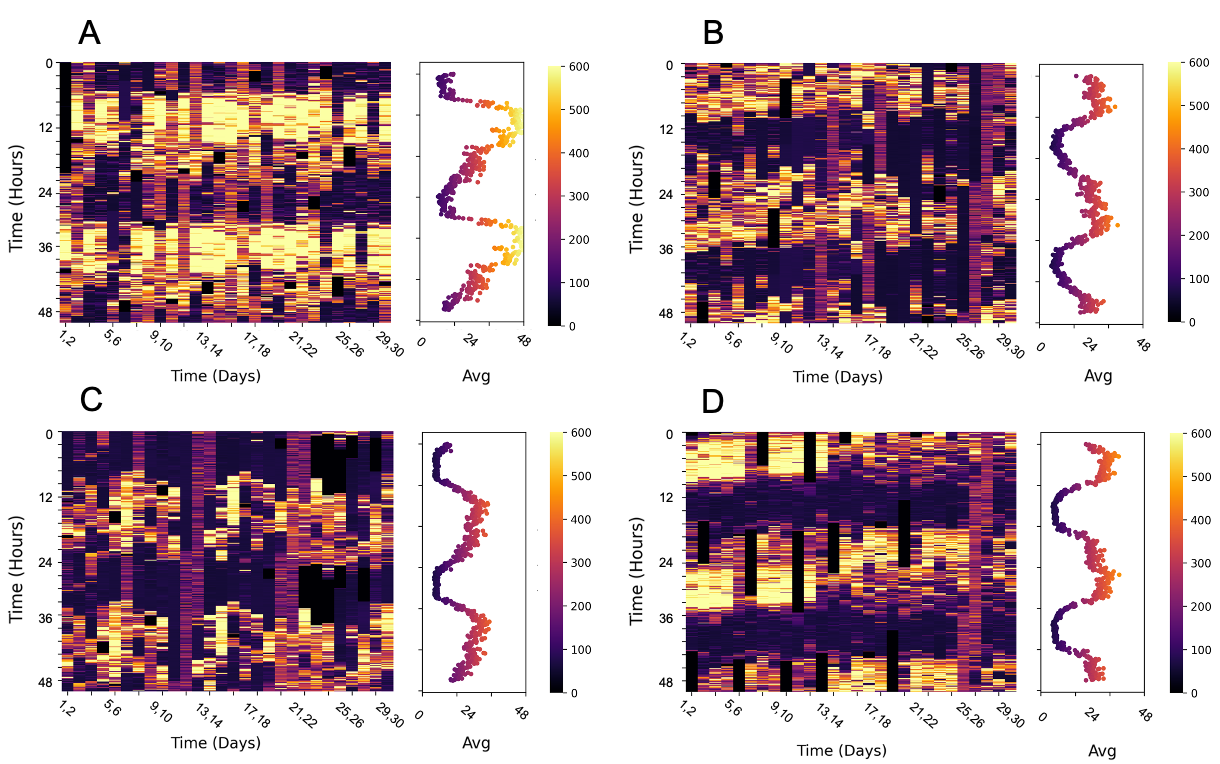}
\caption{\textbf{Double plots of actigraphy data from four patients, illustrating movement intensity over time.} The x-axis represents the first 30 days of the study, while the y-axis spans 48 hours per column, with each column overlapping two consecutive days (e.g., days 1–2, 2–3, etc.) to emphasize daily and circadian activity patterns. Color intensity reflects movement levels, with lighter shades indicating higher activity. The panel to the right of each heatmap displays the average hourly activity across the 30-day study period, providing a concise summary of each patient's circadian rhythm.}
\label{fig:DoublePlot}
\end{figure*}

The wearable device incorporates a tri-axial accelerometer that continuously measures body motion at a sampling frequency of approximately 30--32~Hz. Raw acceleration signals were processed into Activity Counts (AC), one of the most widely used summary measures in actigraphy research for quantifying physical activity intensity and sleep--wake behavior \cite{AncoliIsrael2003,Martin2011}. Briefly, the maximum acceleration observed during each one-second interval was identified and aggregated over consecutive 30-second epochs to generate a single activity count representing movement intensity during that epoch. The resulting longitudinal activity count time series provides a continuous representation of each participant's behavioral activity across both daytime and nighttime periods. Representative examples of these recordings are shown in Figure~\ref{fig:DoublePlot}, where double-plotted actigraphy heatmaps illustrate day-to-day activity patterns and their corresponding circadian profiles.

These high-temporal-resolution activity recordings form the basis of all subsequent analyses. From the activity count time series, we extracted two complementary groups of wearable-derived behavioral features. The first group comprises conventional activity-based measures that summarize movement intensity and short-term temporal patterns, whereas the second group consists of circadian-derived behavioral features that characterize the temporal organization, regularity, and stability of the rest--activity cycle. The feature extraction procedures for both groups are described in the following section.

\subsection{Feature extraction}
\label{sec:data}
\subsubsection{Activity Count Features}
The raw actigraphy recordings were represented as longitudinal Activity Count time series sampled at 30-second intervals, as described in the previous section. To facilitate feature extraction while reducing high-frequency variability, the activity count time series was partitioned into consecutive, non-overlapping 5-minute windows, with each window containing ten 30-second activity count measurements. This segmentation provided a standardized temporal representation of participants' behavioral activity while preserving the overall longitudinal structure of the recordings.

Because continuous wearable monitoring inevitably contains periods of missing or invalid measurements resulting from temporary device removal, signal loss, or non-compliance, missing observations were imputed using a bootstrap-based resampling procedure prior to feature extraction. This preprocessing step reduced the influence of missing data while maintaining the statistical characteristics of the observed activity patterns.

Following preprocessing, a comprehensive set of conventional actigraphy features was extracted from each 5-minute window to characterize participants' behavioral activity. These features summarize multiple aspects of physical movement, including overall activity intensity, temporal variability, distributional characteristics, and short-term temporal dynamics. The resulting feature vectors served as the conventional actigraphy representation used for subsequent machine learning analyses.

\subsubsection{Circadian Rhythm Features}

To characterize participants' daily rest--activity rhythms, circadian features were extracted from the longitudinal activity count time series using cosinor analysis. Cosinor analysis is a parametric approach commonly used to model approximately periodic biological rhythms by fitting a cosine function with a prespecified period to the observed measurements. In the present study, a 24-hour period was assumed to represent the dominant circadian cycle in daily activity:

\begin{equation}
Y(t)=M+K\cos\left(\frac{2\pi t}{\tau}+\phi\right),
\end{equation}

where $Y(t)$ denotes the estimated activity count at time $t$, $\tau$ is the assumed circadian period and was fixed at 24 hours, $M$ is the mesor, $K$ is the amplitude, and $\phi$ is the acrophase. Together, these parameters provide a compact representation of the level, strength, and timing of an individual's daily activity rhythm.

The \textit{mesor}, or Midline Estimating Statistic of Rhythm, represents the rhythm-adjusted mean around which the fitted 24-hour activity pattern oscillates. Unlike a simple arithmetic mean, the mesor is estimated from the fitted cosine model and therefore reflects the central level of the modeled circadian rhythm. In the context of actigraphy, a higher mesor generally indicates a higher overall level of daily activity, whereas a lower mesor reflects a lower baseline level of movement across the observation period.

The \textit{amplitude} represents the vertical distance between the mesor and the maximum value of the fitted cosine curve. It therefore quantifies the magnitude of variation in activity across the 24-hour cycle. A larger amplitude indicates a more pronounced distinction between periods of higher and lower activity, consistent with a stronger or more clearly expressed daily rest--activity rhythm. In contrast, a smaller amplitude suggests a flatter rhythm, with less separation between active and inactive periods. Because amplitude is measured relative to the mesor, the total modeled difference between the daily peak and trough is equal to twice the amplitude.

The \textit{acrophase} describes the timing of the maximum value of the fitted circadian curve. It indicates when peak activity is expected to occur within the 24-hour cycle and therefore provides information about the temporal alignment of an individual's behavioral rhythm. Earlier acrophase values correspond to earlier peak activity timing, whereas later values indicate a shift toward later peak activity. Depending on the implementation, acrophase may be expressed as an angular quantity in radians or degrees and subsequently converted into clock time for interpretation. In the present analysis, acrophase was interpreted relative to the beginning of the 24-hour observation cycle.

These three parameters capture complementary aspects of circadian behavior. The mesor summarizes the overall activity level, the amplitude reflects the strength or prominence of the daily oscillation, and the acrophase characterizes the timing of peak activity. Considered jointly, they provide a concise description of the participant's rest--activity organization that is not captured by activity magnitude alone. The resulting mesor, amplitude, and acrophase estimates were used as circadian-derived predictors in the subsequent machine learning analyses for differentiating ISI-defined insomnia categories.

\subsection{Machine learning approach}
\label{sec:data}
To classify participants according to the presence or absence of each insomnia category, we employed logistic regression classifiers. Logistic regression was selected because the classification task involved a binary clinical outcome and a low-dimensional feature space. Specifically, models based on conventional actigraphy used a single predictor, whereas circadian-based models incorporated three predictors (mesor, amplitude, and acrophase). Given this modest number of explanatory variables, logistic regression provides an appropriate balance between predictive performance, statistical robustness, and interpretability. Furthermore, the relatively large study cohort (approximately 2,300 participants) supports stable estimation of model parameters while minimizing the risk of overfitting. As a generalized linear model, logistic regression also enables straightforward interpretation of the relationship between wearable-derived behavioral features and the probability of belonging to a given insomnia category.

\subsection{Insomnia Severity Measures}

The AURORA dataset includes longitudinal assessments of insomnia severity collected at six study time points: pre-trauma (PRE), Week 2 (WK2), Week 8 (WK8), Month 3 (M3), Month 6 (M6), and Month 12 (M12). Insomnia symptoms were assessed using a modified version of the Insomnia Severity Index (ISI), a validated seven-item questionnaire that evaluates multiple dimensions of insomnia, including difficulty falling asleep, difficulty staying asleep, early morning awakening, satisfaction with current sleep patterns, the extent to which sleep problems are noticeable to others, distress caused by sleep difficulties, and the impact of sleep problems on daily functioning.

Each ISI item is rated on a five-point Likert scale ranging from 0 to 4, with higher scores indicating greater symptom severity. The item scores are summed to obtain a total ISI score ranging from 0 to 28:

\begin{equation}
\text{ISI Score} = \sum_{i=1}^{7} \text{ISI}_i
\end{equation}

Higher total scores indicate greater insomnia severity.

For the pre-trauma assessment (PRE), participants retrospectively reported their sleep experiences during the 30 days preceding the traumatic event. Consequently, the PRE assessment reflects recalled pre-trauma sleep characteristics rather than prospectively collected baseline measurements. Follow-up assessments evaluated insomnia symptoms over predefined reference periods, including the previous two weeks at the Week 2 visit and the previous 30 days at the Month 3, Month 6, and Month 12 follow-up visits. At each assessment, the dataset provides both a continuous ISI total score and a corresponding categorical insomnia severity classification, enabling analyses using either continuous or categorical outcome measures.

\subsection{ISI Severity Categories}

The continuous ISI raw score was converted into four clinically interpretable severity categories following the scoring scheme provided in the AURORA data dictionary.

\begin{table}[tb]
\caption{Classification of Insomnia Severity Index (ISI) scores into insomnia severity categories.}
\centering
\footnotesize

\begin{tabular}{|l|c|c|}
\noalign{\hrule height 1.2pt}

\textbf{Insomnia Severity}
& \textbf{Class}
& \textbf{ISI Score} \\

\noalign{\hrule height 0.9pt}

No clinically significant insomnia
& 0
& 0--7 \\\hline

Subthreshold insomnia
& 1
& 8--14 \\\hline

Moderate clinical insomnia
& 2
& 15--21 \\\hline

Severe clinical insomnia
& 3
& 22--28 \\

\noalign{\hrule height 1.2pt}
\end{tabular}

\label{tab:isi_categories}
\end{table}

For binary classification analyses, participants with ISI raw scores of 15 or greater (categories 2 and 3) were classified as having clinical insomnia, whereas participants with ISI raw scores below 15 (categories 0 and 1) were classified as not having clinical insomnia.

\subsection{Participant Distribution Across Insomnia Severity Categories}

Table~\ref{tab:isi_distribution} summarizes participant availability and the distribution of insomnia severity categories across all study time points. The clinical insomnia category represents the combined prevalence of moderate and severe clinical insomnia (ISI $\geq 15$).

\begin{table}[tb]
\caption{Distribution of participants across Insomnia Severity Index (ISI) severity categories at each assessment time point. Values are reported as $n$ (\%). Percentages are calculated relative to the number of participants available at each assessment.}
\centering

\resizebox{\textwidth}{!}{
\begin{tabular}{|l|c|c|c|c|c|c|}
\noalign{\hrule height 1.2pt}

\multirow{2}{*}{\textbf{ISI Severity}}
& \textbf{PRE}
& \textbf{WK2}
& \textbf{WK8}
& \textbf{M3}
& \textbf{M6}
& \textbf{M12} \\
\cline{2-7}

& \textbf{($n$=2,278)}
& \textbf{($n$=1,799)}
& \textbf{($n$=1,663)}
& \textbf{($n$=1,589)}
& \textbf{($n$=1,357)}
& \textbf{($n$=1,075)} \\

\noalign{\hrule height 0.9pt}

No clinically significant insomnia (0--7)
& 1,149 (50.4)
& 422 (23.5)
& 450 (27.1)
& 512 (32.2)
& 478 (35.2)
& 431 (40.1)
\\\hline

Subthreshold insomnia (8--14)
& 620 (27.2)
& 573 (31.9)
& 548 (33.0)
& 474 (29.8)
& 410 (30.2)
& 329 (30.6)
\\\hline

Moderate clinical insomnia (15--21)
& 382 (16.8)
& 530 (29.5)
& 426 (25.6)
& 383 (24.1)
& 323 (23.8)
& 214 (19.9)
\\\hline

Severe clinical insomnia (22--28)
& 127 (5.6)
& 274 (15.2)
& 239 (14.4)
& 220 (13.8)
& 146 (10.8)
& 101 (9.4)
\\

\noalign{\hrule height 1.2pt}
\end{tabular}
}
\label{tab:isi_distribution}
\end{table}

The dataset provides longitudinal information on insomnia severity from recalled pre-trauma sleep status through 12 months of follow-up. Participant availability decreased over time, from 2,278 individuals at the pre-trauma assessment to 1,075 individuals at Month 12. The prevalence of clinical insomnia, defined as moderate or severe insomnia, increased substantially following trauma exposure, rising from 22.3\% at the pre-trauma assessment to 44.7\% at Week 2. Clinical insomnia prevalence subsequently declined throughout follow-up but remained elevated relative to pre-trauma levels through Month 12, where 29.3\% of participants met criteria for clinical insomnia. These data support both multi-class prediction of insomnia severity categories and binary classification of clinical versus non-clinical insomnia.


\section{Results}
\label{sec:results}

Table~\ref{tab:isi_classification_results} summarizes the classification performance obtained using Activity Count and Circadian feature sets for differentiating each ISI category across the six longitudinal assessment periods. A clear temporal trend was observed, with classification performance generally improving as progressively longer durations of wearable monitoring became available. Across both feature sets, the largest improvements in AUROC occurred between the early follow-up assessments (PRE and WK2) and the later assessment periods (WK8--M12), indicating that longer-term behavioral recordings provide substantially more informative representations of insomnia severity than short-term monitoring.

Importantly, the results suggest that approximately eight weeks of wearable monitoring represent the earliest practical time point at which meaningful discrimination begins to emerge across all ISI categories. At WK8, all four binary classification tasks achieved AUROC values approaching or exceeding 0.60, with performance continuing to improve modestly through the M3, M6, and M12 assessments. In contrast, models developed using only baseline or two-week recordings generally produced AUROC values close to chance, indicating that these shorter monitoring intervals provide insufficient longitudinal behavioral information for reliable classification.

The duration of monitoring influenced ISI categories differently. Participants without clinically significant insomnia were consistently the easiest to identify, achieving the highest discrimination throughout follow-up and reaching a maximum AUROC of 0.693 at M12 using Circadian features. Moderate insomnia showed a gradual improvement over time, with AUROC increasing from approximately 0.50 at baseline to nearly 0.69 by M12, suggesting that intermediate levels of insomnia require several weeks of behavioral observation before becoming distinguishable. Severe insomnia remained the most challenging category because of its relatively low prevalence, although discrimination improved steadily with longer monitoring and reached its highest performance at M6 (AUROC = 0.680). Subthreshold insomnia exhibited intermediate performance, with AUROC values increasing from approximately 0.50 at baseline to 0.67 by M12.

Comparison of the two feature representations showed that Circadian features consistently achieved performance comparable to or slightly better than conventional Activity Count features, particularly after eight weeks of monitoring. Although the absolute improvements were generally modest, the results suggest that circadian characteristics capture complementary information regarding the temporal organization of daily behavior beyond overall activity magnitude. Collectively, these findings demonstrate that the duration of longitudinal wearable monitoring is a critical determinant of insomnia classification performance and that approximately eight weeks of monitoring provide the earliest practical interval for obtaining clinically meaningful discrimination across the full spectrum of ISI-defined insomnia severity.

\begin{table}[htbp]
\centering
\caption{Classification performance of Activity Count (AC) and Circadian (Circ) feature sets for binary prediction of Insomnia Severity Index (ISI) categories across six assessment time points (PRE, WK2, WK8, M3, M6, and M12). Performance is reported using Accuracy, F1-score, Precision, Recall, and ROC-AUC. Circadian-derived features generally achieved comparable or superior performance to Activity Count features, with the greatest improvements observed at later follow-up assessments.}
\label{tab:isi_classification_results}
\resizebox{\textwidth}{!}{
\begin{tabular}{llcccccccccc}
\toprule
\multirow{2}{*}{\textbf{Timepoint}} &
\multirow{2}{*}{\textbf{ISI Label}} &
\multicolumn{2}{c}{\textbf{Accuracy}} &
\multicolumn{2}{c}{\textbf{F1}} &
\multicolumn{2}{c}{\textbf{Precision}} &
\multicolumn{2}{c}{\textbf{Recall}} &
\multicolumn{2}{c}{\textbf{AUROC}} \\
\cmidrule(lr){3-4}\cmidrule(lr){5-6}\cmidrule(lr){7-8}\cmidrule(lr){9-10}\cmidrule(lr){11-12}
&& AC & Circ & AC & Circ & AC & Circ & AC & Circ & AC & Circ\\
\midrule

\multirow{4}{*}{PRE}
& No clinically significant insomnia & 0.516 & 0.518 & 0.554 & 0.541 & 0.513 & 0.515 & 0.602 & 0.570 & 0.511 & 0.535 \\
& Subthreshold insomnia & 0.527 & 0.543 & 0.349 & 0.330 & 0.277 & 0.273 & 0.473 & 0.419 & 0.505 & 0.505 \\
& Clinical insomnia (moderate) & 0.530 & 0.504 & 0.257 & 0.246 & 0.174 & 0.164 & 0.492 & 0.487 & 0.516 & 0.495 \\
& Clinical insomnia (severe) & 0.613 & 0.548 & 0.077 & 0.085 & 0.044 & 0.048 & 0.291 & 0.386 & 0.434 & 0.434 \\

\midrule

\multirow{4}{*}{WK2}
& No clinically significant insomnia & 0.477 & 0.505 & 0.305 & 0.304 & 0.202 & 0.205 & 0.628 & 0.590 & 0.544 & 0.554 \\
& Subthreshold insomnia & 0.485 & 0.492 & 0.372 & 0.371 & 0.267 & 0.268 & 0.613 & 0.604 & 0.542 & 0.528 \\
& Clinical insomnia (moderate) & 0.448 & 0.458 & 0.352 & 0.346 & 0.241 & 0.240 & 0.655 & 0.623 & 0.530 & 0.533 \\
& Clinical insomnia (severe) & 0.441 & 0.510 & 0.201 & 0.208 & 0.121 & 0.129 & 0.592 & 0.540 & 0.529 & 0.534 \\

\midrule

\multirow{4}{*}{WK8}
& No clinically significant insomnia & 0.500 & 0.471 & 0.357 & 0.339 & 0.239 & 0.224 & 0.711 & 0.696 & 0.612 & 0.592 \\
& Subthreshold insomnia & 0.456 & 0.500 & 0.420 & 0.410 & 0.281 & 0.285 & 0.830 & 0.732 & 0.576 & 0.599 \\
& Clinical insomnia (moderate) & 0.422 & 0.490 & 0.337 & 0.344 & 0.214 & 0.226 & 0.794 & 0.726 & 0.574 & 0.607 \\
& Clinical insomnia (severe) & 0.452 & 0.486 & 0.211 & 0.216 & 0.124 & 0.129 & 0.707 & 0.686 & 0.602 & 0.597 \\

\midrule

\multirow{4}{*}{M3}
& No clinically significant insomnia & 0.504 & 0.508 & 0.416 & 0.405 & 0.282 & 0.277 & 0.797 & 0.754 & 0.642 & 0.621 \\
& Subthreshold insomnia & 0.464 & 0.512 & 0.394 & 0.381 & 0.257 & 0.258 & 0.846 & 0.730 & 0.621 & 0.636 \\
& Clinical insomnia (moderate) & 0.447 & 0.495 & 0.327 & 0.323 & 0.205 & 0.208 & 0.809 & 0.726 & 0.607 & 0.611 \\
& Clinical insomnia (severe) & 0.450 & 0.531 & 0.204 & 0.212 & 0.118 & 0.126 & 0.736 & 0.664 & 0.599 & 0.608 \\

\midrule

\multirow{4}{*}{M6}
& No clinically significant insomnia & 0.548 & 0.580 & 0.423 & 0.420 & 0.287 & 0.294 & 0.799 & 0.734 & 0.669 & 0.668 \\
& Subthreshold insomnia & 0.519 & 0.538 & 0.378 & 0.354 & 0.245 & 0.236 & 0.820 & 0.712 & 0.642 & 0.627 \\
& Clinical insomnia (moderate) & 0.508 & 0.562 & 0.316 & 0.313 & 0.196 & 0.201 & 0.814 & 0.715 & 0.667 & 0.658 \\
& Clinical insomnia (severe) & 0.464 & 0.575 & 0.161 & 0.175 & 0.089 & 0.100 & 0.814 & 0.711 & 0.645 & 0.680 \\

\midrule

\multirow{4}{*}{M12}
& No clinically significant insomnia & 0.608 & 0.632 & 0.406 & 0.400 & 0.283 & 0.288 & 0.719 & 0.659 & 0.686 & 0.693 \\
& Subthreshold insomnia & 0.608 & 0.633 & 0.352 & 0.337 & 0.231 & 0.227 & 0.748 & 0.656 & 0.682 & 0.674 \\
& Clinical insomnia (moderate) & 0.599 & 0.624 & 0.255 & 0.253 & 0.154 & 0.155 & 0.739 & 0.687 & 0.694 & 0.688 \\
& Clinical insomnia (severe) & 0.577 & 0.600 & 0.118 & 0.103 & 0.065 & 0.057 & 0.655 & 0.526 & 0.641 & 0.585 \\

\bottomrule
\end{tabular}}
\end{table}

\begin{figure*}
\centering
\includegraphics[width=0.9\textwidth]{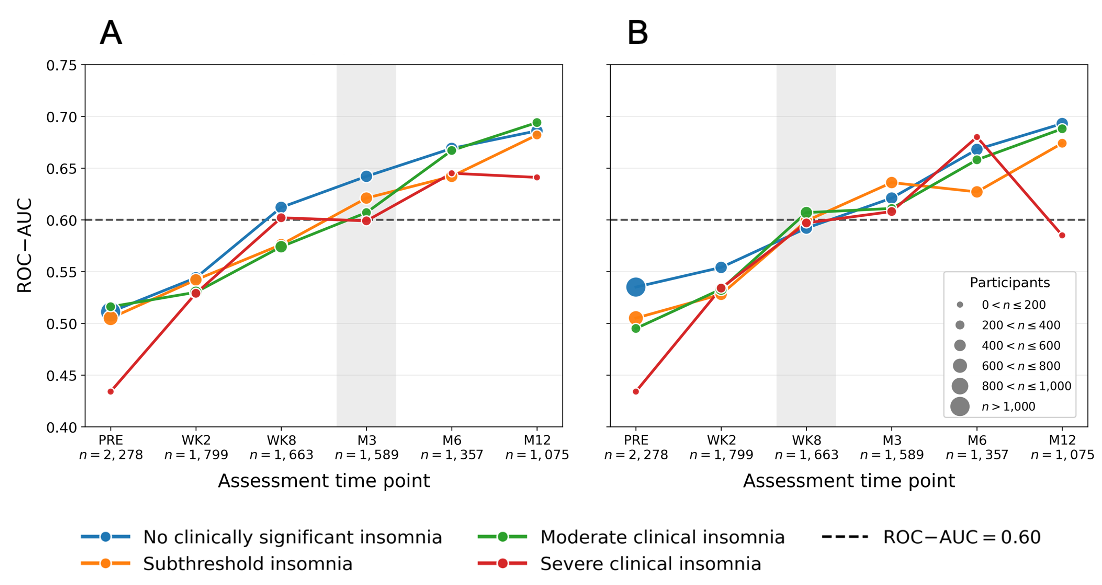}
\caption{ROC–AUC values for predicting insomnia severity at successive follow-up assessments using (A) actigraphy-derived features and (B) circadian rhythm features. Results are shown separately for no clinically significant, subthreshold, moderate clinical, and severe clinical insomnia. Marker size reflects the number of participants available at each assessment, with sample sizes listed below the x-axis. The dashed line denotes a ROC–AUC of 0.60, and the shaded regions indicate the earliest assessment at which predictive performance exceeded this threshold for most insomnia severity categories (WK8 for circadian features and M3 for actigraphy features).}
\label{fig:predictive_performance_over_time}
\end{figure*}

Figure~\ref{fig:predictive_performance_over_time} provides a complementary visualization of how classification performance evolves as progressively longer durations of wearable monitoring become available. Across both feature representations, AUROC values increased steadily over time, demonstrating that longitudinal behavioral recordings become increasingly informative for differentiating insomnia severity. The greatest improvements were observed during the first eight weeks of monitoring, whereas subsequent follow-up periods (M3--M12) yielded comparatively smaller gains, suggesting diminishing returns from extending the monitoring duration beyond approximately two months. This pattern indicates that most of the clinically relevant behavioral information required for insomnia classification is accumulated during the early stages of longitudinal monitoring.

A key finding illustrated in Figure~\ref{fig:predictive_performance_over_time} is that approximately eight weeks of wearable monitoring represents the earliest practical interval at which reliable discrimination begins to emerge across the full spectrum of ISI-defined insomnia categories. For the circadian feature set (Figure~\ref{fig:predictive_performance_over_time}B), all four classification tasks reached or approached an AUROC of 0.60 by the WK8 assessment, whereas the Activity Count models (Figure~\ref{fig:predictive_performance_over_time}A) generally required until the M3 assessment to consistently achieve a comparable level of discrimination. This finding suggests that circadian-derived behavioral features capture meaningful alterations in daily behavioral organization earlier than conventional activity-based measures.

The figure also highlights important differences among insomnia severity categories. Participants without clinically significant insomnia consistently achieved the highest ROC--AUC values throughout follow-up, indicating that this group exhibited the most distinctive behavioral patterns. In contrast, severe clinical insomnia remained the most challenging category to classify, likely reflecting its lower prevalence and greater heterogeneity, although discrimination improved substantially with increasing monitoring duration. Subthreshold and moderate clinical insomnia demonstrated intermediate behavior, with performance improving steadily over time and approaching that of the no-insomnia group after extended monitoring.

Finally, comparison of the two panels indicates that circadian-derived features produced more consistent classification performance across all insomnia categories than conventional Activity Count features. Although the absolute improvements in ROC--AUC were generally modest, the circadian models exhibited less variability across ISI categories and reached clinically meaningful discrimination earlier, supporting the hypothesis that the temporal organization of daily activity provides complementary information beyond overall activity magnitude. Collectively, these findings reinforce that monitoring duration is a critical determinant of wearable-based insomnia classification performance and demonstrate that approximately eight weeks of longitudinal wearable monitoring provides a practical minimum observation period for achieving reliable differentiation of insomnia severity.

\section{Discussion}
\label{sec:discussion}

This study investigated how the duration of longitudinal wearable monitoring influences the ability to objectively differentiate clinically meaningful levels of insomnia severity using wearable-derived behavioral features. Unlike previous studies that evaluated wearable-based models using fixed monitoring periods, our work systematically quantified how classification performance evolves as progressively longer durations of behavioral data become available. Importantly, the relationship between monitoring duration and classification performance is not necessarily monotonic. Although longer observation periods have the potential to capture more stable behavioral patterns, additional data do not always translate into improved predictive performance if they contribute little new information or introduce additional behavioral variability. For example, our recent work on traumatic brain injury prediction using longitudinal sleep--wake data demonstrated that extending the monitoring duration beyond an optimal interval did not further improve predictive performance because the additional observations primarily contributed redundant or noisy information rather than new predictive signal \cite{Saghafi2026}. In contrast, the present study demonstrates that insomnia severity benefits from substantially longer behavioral observation, with approximately eight weeks of monitoring representing the earliest practical duration at which reliable discrimination emerges across all ISI-defined categories.

The observed improvement in classification performance with increasing monitoring duration is consistent with the biological and behavioral characteristics of insomnia. Unlike acute physiological measurements, wearable devices capture daily behavioral manifestations of sleep disruption that are inherently variable from one day to the next. Individual nights are influenced by numerous transient factors, including stress, work schedules, social obligations, illness, and environmental conditions. Consequently, short monitoring periods may predominantly reflect temporary behavioral fluctuations rather than an individual's underlying sleep phenotype. Longer observation periods allow these transient effects to average out while revealing more stable behavioral patterns that are more representative of habitual sleep--wake regulation. This observation aligns with the growing concept of digital phenotyping, in which longitudinal behavioral measurements provide richer and more clinically meaningful information than isolated observations by characterizing persistent behavioral traits rather than short-term variability \cite{DEZAMBOTTI2019,Depner2019}.

An important finding of this study is that the minimum monitoring duration required for meaningful classification differed according to insomnia severity. Participants without clinically significant insomnia exhibited distinctive behavioral patterns that could be identified relatively early and consistently achieved the highest classification performance throughout follow-up. In contrast, moderate and severe insomnia required substantially longer observation periods before reliable discrimination became possible. This finding suggests that progressively more severe insomnia is not simply associated with stronger behavioral signatures, but rather with greater heterogeneity in daily behavioral organization. Individuals with moderate or severe insomnia often differ substantially with respect to symptom presentation, coping strategies, medication use, occupational demands, treatment history, and comorbid psychiatric conditions, all of which may influence activity patterns measured by wearable devices. Consequently, behavioral signatures associated with severe insomnia may only become distinguishable after sufficient longitudinal observation captures persistent disruptions in daily activity organization.

The comparison between conventional Activity Count and circadian-derived behavioral features further illustrates the importance of preserving temporal information in wearable recordings. Activity Count primarily summarizes movement intensity, whereas circadian features quantify the temporal organization, regularity, and phase of the rest--activity rhythm. Although both feature representations produced comparable overall performance, circadian-derived features consistently achieved similar or slightly higher discrimination and reached clinically meaningful performance earlier than Activity Count alone. These findings support previous evidence that circadian organization constitutes an important dimension of sleep behavior that is not fully captured by activity magnitude alone \cite{AncoliIsrael2003,Zee2007,Depner2019}. Rather than replacing conventional actigraphy measures, circadian-derived behavioral features appear to provide complementary information that may improve objective assessment of insomnia.

From a clinical perspective, our findings have important implications for the design of wearable-based monitoring protocols. Many previous studies have adopted arbitrary monitoring durations ranging from several days to multiple weeks without explicitly evaluating whether these observation periods are sufficient for reliable assessment. Our results indicate that monitoring duration should be considered an integral component of wearable study design rather than merely a logistical consideration. Specifically, approximately eight weeks of wearable monitoring appears to provide a practical balance between participant burden and classification performance, whereas extending monitoring beyond this period yields progressively smaller improvements in discrimination. This information may assist clinicians and investigators in selecting monitoring durations that maximize diagnostic utility while minimizing unnecessary data collection and participant burden.

Several limitations should be acknowledged. First, insomnia severity was determined using self-reported ISI scores rather than polysomnography or clinician-confirmed diagnoses. Although the ISI is one of the most widely validated clinical instruments for assessing insomnia severity, subjective symptom reports do not always correspond directly to objective behavioral measurements. Second, the present study focused exclusively on wearable-derived behavioral features. Incorporating additional physiological signals, such as heart rate variability, photoplethysmography, skin temperature, or smartphone-derived digital phenotypes, may further improve classification performance. Third, the analyses evaluated each assessment period independently rather than explicitly modeling longitudinal trajectories of insomnia severity. Future work should investigate temporal machine learning models capable of jointly learning behavioral evolution across repeated assessments while predicting future changes in insomnia severity.

In summary, this study demonstrates that the duration of longitudinal wearable monitoring is a critical determinant of objective insomnia classification. Rather than treating wearable recordings as isolated observations, our findings emphasize the importance of longitudinal behavioral monitoring for capturing stable sleep-related behavioral patterns. By identifying approximately eight weeks as the earliest practical monitoring interval for achieving reliable discrimination across ISI-defined insomnia categories, this work provides evidence-based guidance for the design of future wearable studies and highlights the potential of circadian-derived behavioral features as scalable digital biomarkers for objective insomnia assessment.


\section{Conclusions}
\label{sec:conclusions}
This study systematically investigated how the duration of longitudinal wearable monitoring influences the objective classification of insomnia severity using wearable-derived behavioral features. Our findings demonstrate that monitoring duration is a critical determinant of classification performance, with approximately eight weeks of wearable monitoring representing the earliest practical interval at which reliable discrimination emerges across all ISI-defined insomnia categories. Although performance continued to improve with longer observation periods, the greatest gains occurred during the first eight weeks, with only modest improvements thereafter.

We further showed that the amount of longitudinal monitoring required depends on insomnia severity. Participants without clinically significant insomnia could be differentiated relatively early, whereas moderate and severe insomnia required substantially longer observation to achieve meaningful discrimination. In addition, circadian-derived behavioral features consistently achieved performance comparable to, and in several clinically relevant settings better than, conventional Activity Count features, suggesting that the temporal organization of daily activity provides complementary information beyond activity magnitude alone.

Overall, these findings demonstrate that successful wearable-based insomnia assessment depends not only on the behavioral features extracted from wearable devices but also on collecting a sufficient duration of longitudinal observations. By providing evidence-based guidance on the minimum monitoring duration required for reliable classification, this work offers practical insights for the design of future wearable studies, remote patient monitoring protocols, and digital health applications. Future research should investigate multimodal physiological biomarkers and personalized longitudinal modeling approaches to further improve objective insomnia assessment.


\section*{Acknowledgments}
\label{sec:acknowledgments}
 The investigators wish to thank the trauma survivors participating in the studies. Their time and effort during a challenging period of their lives make our efforts to improve recovery for future trauma survivors possible. This material is based upon work supported by the United States Army Medical Research Acquisition Activity (USAMRAA) under Contract No. W81XWH22C0122. The data collection through the AURORA study was supported by the NIMH under U01MH110925, One Mind, and The Mayday Fund. Verily Life Sciences and Mindstrong Health provided some of the hardware and software used to perform study assessments. GC is also funded by the NHLBI grant R01 HL161253. This work was also supported by the National1. The content is solely the responsibility of the authors and does not necessarily represent the official views of any of the funders, who had no role in the data analysis, interpretation, or preparation of this manuscript.

\nolinenumbers

\bibliographystyle{unsrt}
\bibliography{references}
\end{document}